\documentclass[12pt,preprint]{aastex}

\slugcomment{Accepted by ApJ}

\def\MO{M_\odot}

\shorttitle{Sgr\,B2(N): A bipolasr outflow and rotating hot core revealed by ALMA}
\shortauthors{A.E.Higuchi et al.}

\begin{document}
\title{Sgr\,B2(N): A bipolar outflow and rotating hot core revealed by ALMA}

\author{Aya E. Higuchi\altaffilmark{1}, Tetsuo Hasegawa\altaffilmark{2}, Kazuya Saigo\altaffilmark{3}, 
Patricio Sanhueza\altaffilmark{2}, James O. Chibueze\altaffilmark{4}}
\email{aya.higuchi.sci@vc.ibaraki.ac.jp}

\altaffiltext{1}{College of Science, Ibaraki University, 2-1-1 Bunkyo, Mito, 310-8512, Japan}
\altaffiltext{2}{National Astronomical Observatory of Japan 2-21-1 Osawa, Mitaka, Tokyo, 181-8588, Japan}
\altaffiltext{3}{Department of Physical Science, Graduate School of Science, Osaka Prefecture University, 1-1 Gakuen-cho, 
Naka-ku, Sakai, Osaka 599-8531, Japan}
\altaffiltext{4}{Department of Physics and Astronomy, Faculty of Physical Sciences, University of Nigeria, 
Carver Building, 1 University Road, Nsukka, Nigeria}

\begin{abstract}
We present the results of SiO ($2 - 1$) and SO$_{2}$ ($12_{4,8} - 13_{3,11}$) line observations of Sgr\,B2(N) made with the 
Atacama Large Millimeter/submillimeter Array (ALMA) at an angular resolution of $\sim$2$\arcsec$. 
Our analysis of the SiO and SO$_{2}$ line emission reveals a bipolar molecular outflow in an east-west direction whose driving source is located at K2. 
In addition, SO$_{2}$ line core shows a north-south velocity gradient most probably indicating a hot core of molecular gas rotating around K2. 
Fractional abundances of SO$_{2}$ and SiO ($X$(SO$_{2}$) and $X$(SiO), respectively) in the outflowing molecular gas are derived from comparisons with the C$^{18}$O emission. 
Assuming an excitation temperature of $100 \pm 50$~K, 
we calculate $X$(SO$_{2}$) = ${2.3^{+2.6}}_{-0.4}$$\times$10$^{-8}$ and $X$(SiO) = ${1.2^{+0.1}}_{-0.1}$$\times$10$^{-9}$.
The outflow from Sgr\,B2(N) K2 is characterized as a young (5$\times$10$^{3}$~yr) and massive ($\sim$2000~$\MO$), 
but moderately collimated ($\sim$60$^{\circ}$) outflow.
We also report a possible detection of the SiO ($v = 2, J = 2 - 1$) maser emission from the position of K2. 
If confirmed, it would make Sgr\,B2(N) the 4$^{\rm th}$ star forming region associated with SiO masers.
\end{abstract}


\section{INTRODUCTION}

High-mass young stellar objects are usually deeply embedded in their parental dense and massive molecular clumps 
(size$\sim$$\,$1~pc, mass$\sim$$\,$100--1000~$\MO$, density $\sim$10$^{4-5}$~$\rm{cm}^{-3}$)
\citep{rid03, lad03, lad10, hig09, hig10, hig13}, obscuring their early formative stages. 
Their formation timescales of $\sim$ 10$^{5}$~yr are short, and they form in distant clusters 
(e.g., Galactic Center) and associations \cite[e.g.,][]{zin07}. 
All these factors limit our understanding of their formation processes.
High angular resolution observations are indispensable in the efforts to unveil the mystery of high-mass star formation. 
The Atacama Large Millimeter/submillimeter Array (ALMA) provides the high sensitivity, angular 
resolution, and dynamic range to improve our understanding of the formation processes of high-mass stars and their parental clumps
\citep{sanc13,sanc14,bel14,guz14, hig14, hig15,zap15,joh15}.

Sagittarius\,B2 (Sgr\,B2) is a complex of H {\sc ii} regions and giant molecular clouds located near the Galactic Center. 
It is known as one of the most luminous massive star-forming regions in our Galaxy, with a total bolometric luminosity of $\sim$10$^{7}$~$\MO$ \citep{lis89,lis90,lis91}. 
Tens of compact/ultracompact H {\sc ii} regions and H$_2$O, OH, and H$_2$CO maser clusters are located at the three centers of star formation aligned in the 
north-south direction,  
i.e., Sgr B2 (N), (M), and (S) \citep{gen76,ben84,gar86,gau90}. 
Sgr B2(N) is associated with the H {\sc ii} region K, which is further resolved into subcomponents K1 -- K6 \citep{gau90,gau95,dep96}. 
Of these, K1, K2 and K3 have drawn particular attention because they are very compact $(0\farcs3 - 0\farcs6)$ \citep{gau95} and are spatially 
coincident with the H$_2$O masers \citep{kob89}.

A dense core of hot molecular gas was initially found in the K1 -- K3 region by interferometric observations of 
NH$_3$ and HC$_3$N line emission \citep{vog87,gau90,lis93}.
The core is also detected in dust continuum emission at millimeter wavelengths. 
The dust emission peaks at the position of K2 with the peak H$_2$ column density of $N_{{\rm H}{_2}}$ $\sim$ $(0.4 - 8)$ 
$\times$ 10$^{25}$ cm$^{-2}$ \citep{car89,lis93,kua96}. 
The inferred mass of the core ranges from $10^4 - 10^5$ $M_\odot$, for a region with a diameter of $0.2 - 0.4$ pc.
Interferometric and single-dish follow up observations in several molecular lines have revealed that the core is particularly rich 
in large saturated molecules, indicating hot core chemistry with fresh material evaporating from dust mantles 
\citep{gol87,kua94,kua96,liu99,bel08,hol03,qin11,bel13,bel14}.

For the kinematic structure of the molecular gas in this region, two possible interpretations have been presented. 
\cite{vog87} was the first to note the NW -- SE velocity gradient in their NH$_3$ observations. 
\cite{lis93} confirmed the velocity structure in their HC$_3$N ($25 - 24$) observations at $\sim$4$\arcsec$ resolution, 
and suggested that it arises from a very energetic bipolar outflow associated with a source embedded in the dense core. 
On the other hand, \cite{hol03} argued, based on their 1\farcs5 resolution observation of the CH$_3$CH$_2$CN ($5_{1,5} - 4_{1,4}$) line, 
that this velocity gradient indicates the rotation of a large edge-on disk that extends in the east-west direction. 
Our understanding of the kinematics remain unsettled between these two contradicting pictures.

In this paper, we analyze the ALMA archival data taken at $\sim$2$\arcsec$ 
resolution and present images of Sgr\,B2(N) in SiO ($2 - 1$) [hereafter SiO] and SO$_{2}$ ($12_{4,8} - 13_{3,11}$) 
[hereafter SO$_{2}$] line emission. At the adopted distance of 7.8~kpc \citep{rei09}, 2$\arcsec$ corresponds to 0.076 pc. 
The SO$_{2}$ molecule is a good tracer of dense and hot gas typical of hot cores \cite[e.g.,][]{cha97,van03,leu07,jim07}, 
while SiO is an excellent tracer of molecular outflows and shocks \cite[e.g.,][]{dow82,bac91,zap09,leu14}.
The goal is to establish a clearer kinematic picture of the Sgr B2(N) core, and to enable further explorations of the processes 
associated with the on-going high-mass stars formation. 
We also report the possible discovery of the SiO ($v = 2, J = 2 - 1$) maser emission from Sgr B2(N).

\section{OBSERVATIONS}

Sgr\,B2(N) was observed with ALMA \citep{hil10} during Early Science Cycle 0 \citep{bel14}.
The observations were done in 1 execution with 21~12-m antennas and 3 executions with 26~12-m antennas at an angular 
resolution of $\sim$2$\arcsec$.
The maximum baseline length achieved during the observations was 440~m.
The full frequency range between 84 and 111~GHz was covered in the spectral scan mode with four overlapping basebands.
The HPBW of the primary beam varies between 74$\arcsec$ at 84 GHz and 56$\arcsec$ at 111 GHz. 
The channel spacing was 244~kHz, and smoothed spectra were 488~kHz ($\sim$1.6~km~s$^{-1}$).

The ALMA calibration includes simultaneous observations of the atmospheric 183~GHz water line with water vapor radiometers. 
The measured water columns in the antenna beam were later used to reduce the atmospheric phase fluctuations. 
Amplitude calibration was done using Neptune.
The quasars B1730-130 and J1700-261 were used to calibrate the bandpass and the complex gain fluctuations, respectively. 
Data reduction was performed using the Common Astronomy Software Applications package (CASA; http://casa.nrao.edu).
Natural Weighting was used for imaging. The sensitivity obtained ranges from 3 to 7 mJy beam$^{-1}$ per channel of 1.6 km s$^{-1}$ 
(see details in Table~\ref{obs}). 

In this paper, we focus on the SO$_{2}$ and SiO lines.
Among several transitions of SO$_{2}$ detected between 84 and 111~GHz in the ALMA observations, 
we select 12$_{4,8}$--13$_{3,11}$ transition for the most part of our analysis because; 
(1) its upper state energy is high ($E_{\rm u} = 110.6$~K), and it traces the hotter and denser region in the immediate vicinity of the young star, and 
(2) this transition is relatively less contaminated by the other coincident lines so that we can 
get clearer kinematic picture of the core.
SiO is an excellent tracer of shocks and molecular outflows \cite[e.g.,][]{dow82,mar97,zap09}.
We also investigate the lines of SiO in the $v$=1,2,3 vibrationally excited states.
Parameters of the data used are summarized in Table \ref{obs}.

\section{RESULTS and DISCUSSION}
\subsection{Line profiles}\label{prof}

Figure \ref{prof} shows the spectra of the SO$_{2}$ and SiO lines integrated within a 20$\arcsec$$\times$20$\arcsec$ 
region centered on the K2 position: ($\alpha_{J2000}$, $\delta_{J2000}$)=[17$^{\mbox h}$47$^{\mbox m}$19$^{\mbox s}.$88, 
~$-$~$\!\!$28$^{\circ}$22$\arcmin$18$\arcsec.$4] \citep{gau95}. 
Many spectral lines other than SO$_{2}$ and SiO are detected in the ALMA data toward Sgr\,B2(N).
We also label the molecular lines near the SO$_{2}$ and SiO lines identified by \cite{bel13} using the IRAM 30~m telescope. 
Although these molecular lines partially overlap with the SiO and SO$_{2}$ lines,
we can separate them because their spatial distributions are different. 

The integrated SO$_{2}$ spectrum in Figure \ref{prof}(a) has an emission peak of 4.0~Jy at 63~km~s$^{-1}$, 
which we adopt for the systemic velocity ($V$$_{\rm{SYS}}$) of Sgr\,B2(N).
We define $\Delta{V}$ as $\Delta{V}$$\equiv$$V$$_{\rm{LSR}}$--$V$$_{\rm{SYS}}$ ($V$$_{\rm{SYS}}$=63~km~s$^{-1}$).
By visual inspection of the SO$_{2}$ spectrum, it is seen a line core from $V_{\rm{LSR}}$ of 56 to 70~km~s$^{-1}$ 
($\Delta{V}$= -7 to +7~km~s$^{-1}$), a blueshifted wing from $V_{\rm{LSR}} \approx 35$ to 56~km~s$^{-1}$ 
($\Delta{V} \approx -28$ to -7~km~s$^{-1}$), 
and a redshifted wing from  $V_{\rm{LSR}} \approx 70$ to 110~km~s$^{-1}$ 
($\Delta{V} \approx +7$ to  +47~km~s$^{-1}$).

The integrated SiO spectrum in Figure \ref{prof}(b) shows a deep absorption from  
$V_{\rm{LSR}} \approx 48$ to 92~km~s$^{-1}$ ($\Delta{V} \approx -15$ to +29~km~s$^{-1}$).
This is consistent with the observations made by \cite{liu98} using the Berkeley-Illinois-Maryland Association (BIMA) array.
In addition, the present SiO spectrum exhibits wing emission 
at blueshifted velocities of $V_{\rm{LSR}} \approx 30$ to 48~km~s$^{-1}$ 
($\Delta{V} \approx -33$ to -15~km~s$^{-1}$) 
and redshifted velocities of $V_{\rm{LSR}} \approx 92$ to 110~km~s$^{-1}$ 
($\Delta{V} \approx +29$ to +47~km~s$^{-1}$).

\subsection{Kinematic structure of SO$_{2}$ and SiO emission}
\subsubsection{SO$_{2}$ line emission}

\begin{figure}
\epsscale{1}
\plotone{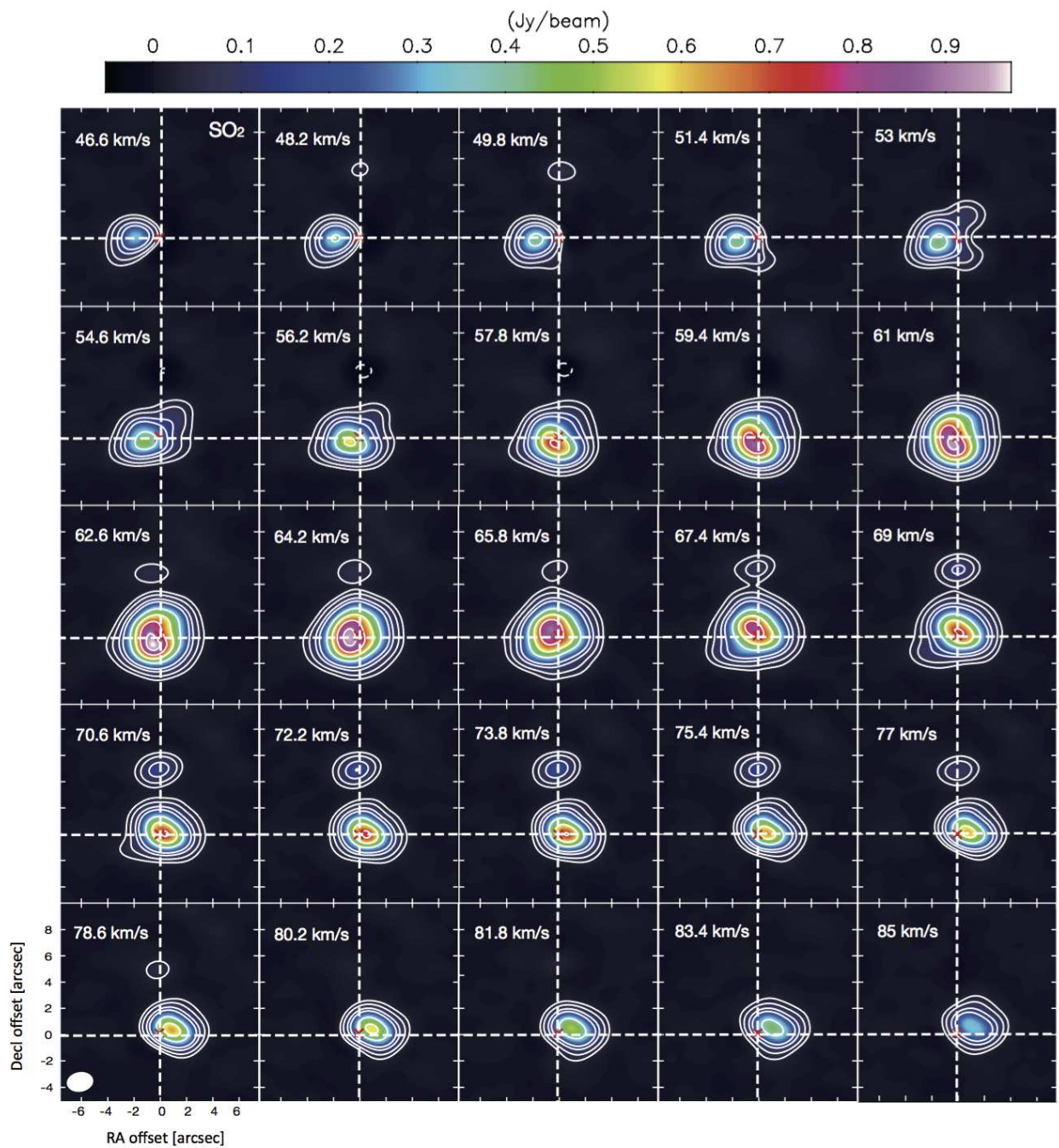}
\caption{
Velocity channel maps of the SO$_{2}$ emission in white contours and color image for Sgr\,B2(N). 
The velocity intervals are 1.6 km s$^{-1}$.
Contours start from -5, 5, 10 to 30~$\sigma$ increasing in intervals of 10~$\sigma$ levels and
then they continue in step of 30~$\sigma$ up to the 150~$\sigma$ level (1~$\sigma$ = 6~mJy~beam$^{-1}$).
Negative contour levels are shown with dashed lines.
The red crosses mark the position of K2 \citep{gau95}.}
\label{so2_ch}
\end{figure}

Figure \ref{so2_ch} shows the channel maps of SO$_{2}$ emission in the velocity range of $V_{\rm{LSR}}$= 47 to 85~km~s$^{-1}$ 
($\Delta{V}$= -16 to +22~km~s$^{-1}$). 
The positions are shown in offsets from the of K2.
At the velocities of the blueshifted wing shown in Figure \ref{so2_ch} 
($V_{\rm{LSR}}$=47 to 56~km~s$^{-1}$, $\Delta{V}$= -16 to -7~km~s$^{-1}$) 
the SO$_{2}$ emission extends to E to SE of K2, 
while at the velocities of the redshifted wing in Figure \ref{so2_ch} 
($V_{\rm{LSR}}$=70 to 85~km~s$^{-1}$, $\Delta{V}$= +7 to +22~km~s$^{-1}$) 
the emission extends to W to NW of K2.
This makes a clear E-W velocity gradient.
However, within the ``line core" ($V_{\rm{LSR}}$=56 to 70~km~s$^{-1}$, also see Figure \ref{prof}), the velocity gradient changes its direction.
At $V_{\rm{LSR}}$= 57.8 to 61~km~s$^{-1}$, the peak of the SO$_{2}$ emission in located to S-SE of K2, 
while at $V_{\rm{LSR}}$= 65.8 to 69 ~km~s$^{-1}$ the peak is NE-N of K2.
This ``line core" component may predominantly originate from the central compact core and is likely tracing a different kinematic structure
\cite[e.g.,][]{sanc14,guz14}.
A secondary peak is seen $\sim$5$\arcsec$ north of K2 at $V_{\rm{LSR}}$$\sim$ 70~km~s$^{-1}$.
This is identified with one of the compact ``quasi-thermal cores", source $h$, observed in CH$_3$OH by \cite{meh97}.

\begin{figure}
\epsscale{1}
\plotone{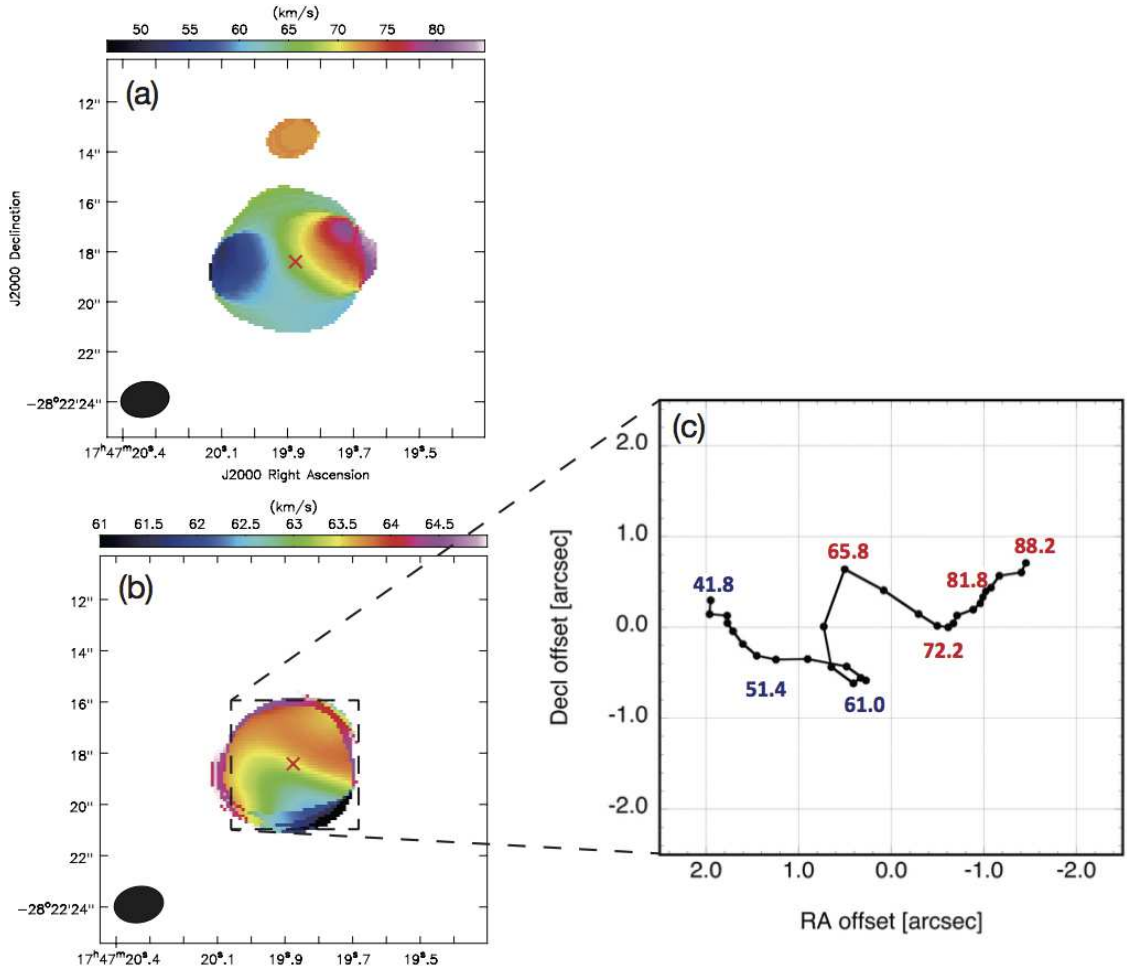}
\caption{(a): The first moment map of SO$_{2}$ emission made using all velocity ranges (33-109~km~s$^{-1}$).
(b): The first moment map of SO$_{2}$ emission made using velocity ranges of 59-65~km~s$^{-1}$.
(c): A zoom into the SO$_{2}$ peak position of Figure \ref{so2_vel}(b).
The red crosses mark the position of K2 \citep{gau95}.}
\label{so2_vel}
\end{figure}

The change of the direction of the velocity gradient described above is clearly shown by the moment maps in Figure \ref{so2_vel}.
The first moment map (intensity-weighted velocity) made from the entire velocity range of $V_{\rm{LSR}}$= 33 to 109~km~s$^{-1}$
($\Delta{V}$= -30 to +46~km~s$^{-1}$) shows the E-W velocity gradient (Figure \ref{so2_vel}(a)).
In contrast, the first moment map made from channels near the systemic velocity ($V_{\rm{LSR}}$=59 to 65~km~s$^{-1}$)
shows the S-N velocity gradient (Figure \ref{so2_vel}(b)).

Figure \ref{so2_vel}(c) shows the positional shift of the SO$_{2}$ emission peak with velocity.
As we noted in Figure \ref{so2_ch}, the peak at the wing velocities ($|\Delta{V}|$ $>$ 7~km~s$^{-1}$) trace the E-W velocity gradient,
while the peak positions at the ``line core" velocities ($|\Delta{V}|$ $<$ 7~km~s$^{-1}$) clearly shows the N-S velocity gradient 
perpendicular to the former.

\subsubsection{SiO line emission}

\begin{figure}
\epsscale{0.8}
\plotone{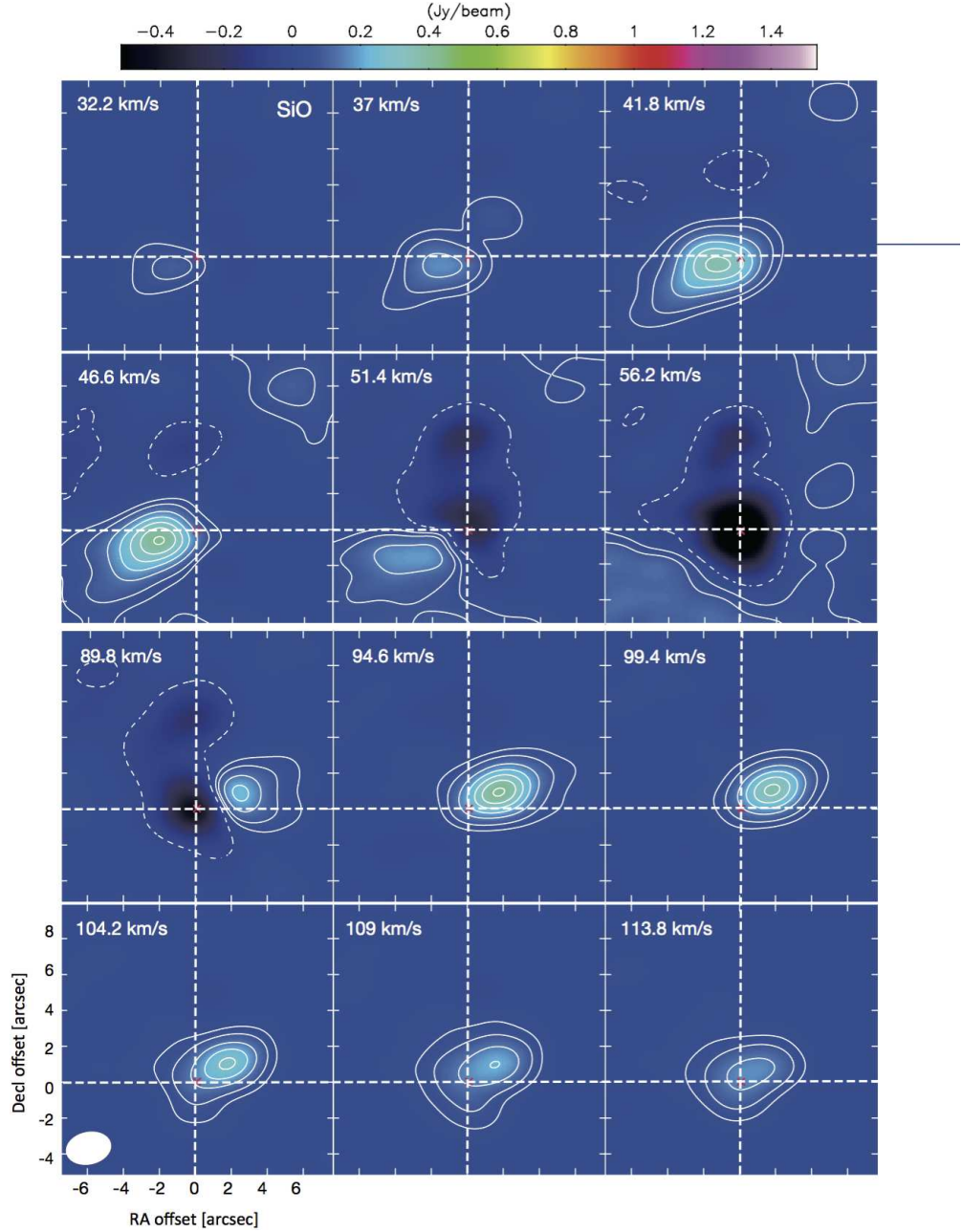}
\caption{Velocity channel maps of the SiO emission in white contours and color image for Sgr\,B2(N). 
Contours start from -5, 5, 10 to 60~$\sigma$ increasing in intervals of 10~$\sigma$ levels 
(1~$\sigma$=7~mJy~beam$^{-1}$). 
Negative contour levels are shown with dashed lines.
The velocity intervals are 4.8 km s$^{-1}$.
The red crosses mark the position of K2.}
\label{sio_ch}
\end{figure}

Figure \ref{sio_ch} shows the velocity channel maps of the SiO wing at emission seen in Figure \ref{prof}(b), i.e., 
blueshifted wing at $V_{\rm{LSR}}$=32 to 51~km~s$^{-1}$ ($\Delta{V}$=-31 to -12~km~s$^{-1}$)
and redshifted wing at $V_{\rm{LSR}}$=90 to 114~km~s$^{-1}$ ($\Delta{V}$= +27 to +51~km~s$^{-1}$). 

The SiO emission at blueshifted wing extends to E to SE of K2, while that at the redshifted wing extends to W to NW of K2.
This is similar to what we see for the SO$_{2}$ emission at the wing velocities, 
and suggests that SiO and SO$_{2}$ emission share the same kinematics.

\subsection{Bipolar outflow and rotating hot core}

As we described in Section~1, the understanding of the kinematic structure of the Sgr\,B2(N) core is unsettled.
Two mutually contradicting interpretations have been put forward so far; 
a) E-W bipolar jet and a disk rotating in N-S direction \cite[e.g.,][]{lis93},
and b) a large disk rotating in E-W direction \citep{hol03}.
What do the new ALMA observations tell us?

\begin{figure}
\rotate
\epsscale{0.8}
\plotone{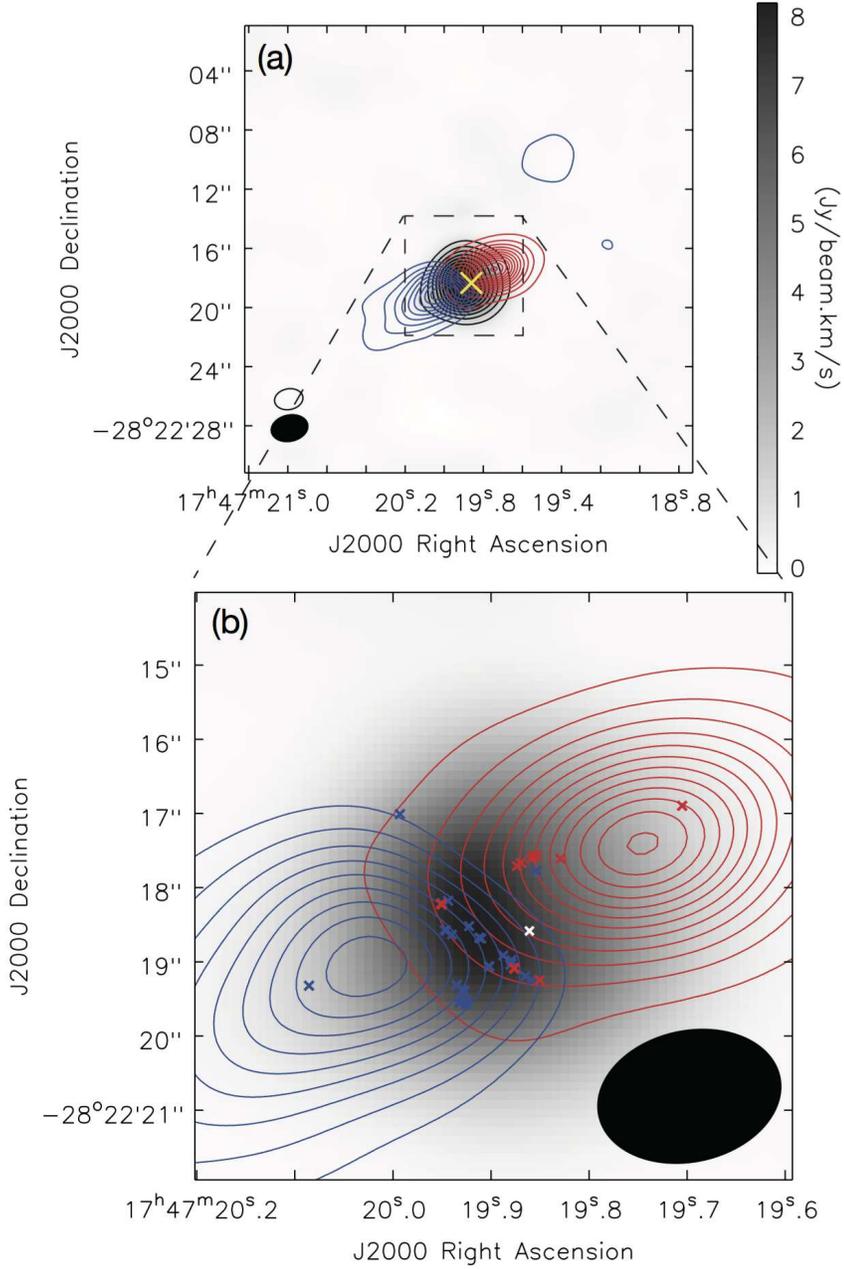}
\caption{
(a) SO$_{2}$ integrated intensity map (gray-scale image and black contours) and the integrated intensity map of the SiO outflows 
(blue and red contours) of Sgr\,B2(N).
The blue color represents blueshifted gas, while the red color represents the redshifted gas. 
The black contours range from 10 to 90$\%$ of the peak emission, in steps of 10$\%$. 
The blue color represents blueshifted gas ($V_{\rm LSR}$ = 29 to 50~km~s$^{-1}$), 
while the red color represents the redshifted gas ($V_{\rm LSR}$ = 91 to 111~km~s$^{-1}$). 
The contours for integrated intensity maps, with intervals of 5$\sigma$, start from the 5$\sigma$ level
(1$\sigma$=0.11~Jy~beam$^{-1}$~km~s$^{-1}$ for both blueshifted and redshifted gas).
The gray-scale bar shows the flux density of the SO$_{2}$ emission.
The yellow cross marks the K2 position.
(b) A zoom into the SO$_{2}$ integrated intensity map and the SiO outflow. 
The blue and red crosses mark the position of the blue- and redshifted water maser spots reported by McGrath et al. (2004).
The white cross marks the position of the possible SiO maser (see Section \ref{maser}). }
\label{outflow}
\end{figure}

The SO$_{2}$ and SiO spectra show the wing emission at high velocities, 
which suggests the presence of an outflow with velocities far exceeding 20~km~s$^{-1}$ 
($\Delta{V} \approx -28$ to +47~km~s$^{-1}$ for SO$_{2}$ and $\Delta{V} \approx -33$ to +47~km~s$^{-1}$ for SiO).
Figure \ref{outflow}(a) shows the spatial distribution of the SiO wing emission that we identify as the bipolar outflow. 
The velocity ranges were selected to separate distinct feature of outflow lobes.
The bipolar outflow lobes are aligned in the E-W direction (P.A.=120$^\circ$$\pm$10$^\circ$), 
and are symmetrically displaced about the peak of the integrated SO$_{2}$ emission near K2.
Figure \ref{outflow}(b) shows the zoom-up of the central region. 
Blue and red crosses mark the position of the H$_2$O maser spots reported by \cite{mcg04} blue- and redshifted with respect to the 
systemic velocity $V_{\rm{SYS}}$ = 63~km~s$^{-1}$, respectively. 
The position and velocity of the masers are consistent with the innermost part of the blueshifted and redshifted outflowing gas traced by SO$_{2}$ and SiO.
The white cross marks the position of the possible SiO maser source presented in Section \ref{maser}.

\begin{figure}
\epsscale{1}
\plotone{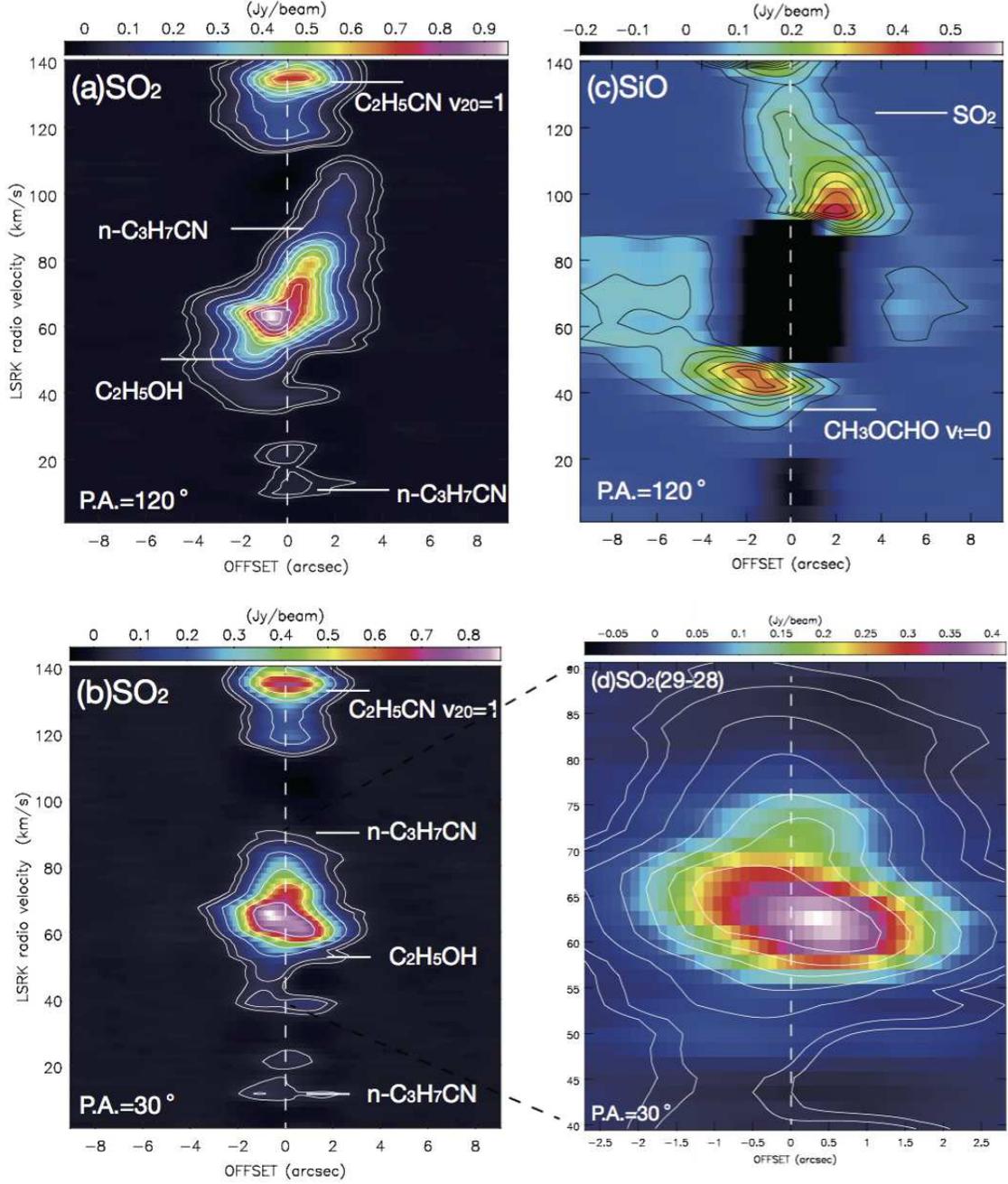}
\caption{Position-velocity diagrams along the outflow in SO$_{2}$ and SiO emission,
P.A.=120$^\circ$ (panel a and c), and perpendicular to the outflow in SO$_{2}$ emission, P.A.=30$^\circ$ (panel b and d). 
(a) contour levels [5, 10, 30, 60, 90, 120, 150, 180, 210, 240, 270, and 300] $\times$ 1$\sigma$ (1~$\sigma$ = 3~mJy~beam$^{-1}$). 
(b) contour levels [5, 10, 30, 60, 90, 120, and 150]  $\times$ $\sigma$ (1~$\sigma$ = 5~mJy~beam$^{-1}$)
(c) contour levels [5, 10, 15, 20, 25, 30, 35, and 40] $\times$ $\sigma$ (1~$\sigma$ = 10~mJy~beam$^{-1}$).
(d)SO$_{2}$(29$_{4,26}$--28$_{5,23}$) (color) and SO$_{2}$(12$_{4,8}$--13$_{3,11}$) (contours), 
contours levels [5, 10, 30, 60, 90, 120, and 150] $\times$ $\sigma$ (1~$\sigma$ = 5~mJy~beam$^{-1}$).
Other molecular lines (e.g., C$_{2}$H$_{5}$OH, n-C$_{3}$H$_{7}$CN) are also displayed.}
\label{pv}
\end{figure}

Figure \ref{pv}(a) shows the observed position-velocity (P-V) diagram of the SO$_{2}$ emission along the major axis of the bipolar outflow 
(P.A.=120$^\circ$). 
The P-V diagram along this axis shows blueshifted and redshifted emission extending to $\Delta{V} \approx -28$ to +47~km~s$^{-1}$.
Figure \ref{pv}(b) shows the observed P-V diagrams of the SO$_{2}$ emission in the direction perpendicular to the axis of the outflow (P.A.=30$^\circ$). 
Between $V$$_{\rm{LSR}}$=59 to 65~km~s$^{-1}$, the peak velocity changes almost linearly as a function of the position with a constant velocity gradient. 
Figure \ref{pv}(c) shows the P-V diagram of the SiO emission along the major axis of the 
blueshifted and redshifted components (P.A.=120$^\circ$).
The blueshifted and redshifted SiO emission is seen extending up to $\Delta{V} \approx -33$ to +47~km~s$^{-1}$. 
This shows that the SiO and SO$_{2}$ lines share the same kinematics except for the severe foreground absorption in SiO at low velocities ($\Delta{V}$= -15 to +29~km~s$^{-1}$).

From the results presented above, we summarize the evidences for the existence of the bipolar outflow as follows;
(1) presence of high-velocity gas (wing components) in the spectra of shock tracers,
(2) the bipolar structure is highly symmetric and well traced by the SiO line, which is a good tracer of outflows,
(3) H$_{2}$O masers are associated with the bipolar structure, and 
(4) the high velocity motion cannot be gravitationally bound by the central core.
In fact, were the high velocity emission a result of a gravitationally bound circular motion, the mass required inside its orbit is estimated by the dynamical mass $M_{\rm{dyn}}$; 
\begin{eqnarray}
\label{dyn}
{M_{\rm{dyn}}} = \frac{\omega^{2}r^{3}}{G} \quad,
\end{eqnarray}
where $\omega$ is an angular velocity, and $r$ is a rotational radius.
From the observed angular velocity of 1.6$\times$10$^{-11}$~s$^{-1}$ and a 0.08~pc radius, 
we derive a dynamical mass of $\sim$3$\times$10$^{4}$~$\MO$ needed to gravitationally bind the motion, which should be seen as a lower limit because we do not correct for the inclination of the motion. 
This value is higher than the mass of the core derived from the 1.3~mm and 1.1~mm continuum emission with larger core radius (9$\times$10$^{3}$~$\MO$ by Lis et al. 1993; 1$\times$10$^{4}$~$\MO$ 
by Qin et al. 2008; 4--20$\times$10$^{3}$~$\MO$ by Liu et al. 1999), and shows that the gas responsible for the high velocity wing emission cannot be gravitationally bound by the central core. 

Meanwhile, we consider that the N-S velocity gradient observed in SO$_{2}$ at P.A. = 30$^\circ$, {\it i.e.,} orthogonal to the bipolar outflow, is a signature of the rotating motion of the core (see Figure \ref{pv}(b)). 
It has been reported that SO$_{2}$ traces rotating cores in other regions of massive star formation \cite[e.g.,][]{belt14,guz14}. 
From Figure \ref{pv}(b), we determine a velocity gradient of 88~km~s$^{-1}$~pc$^{-1}$ or an angular velocity $\omega = 2.8 \times$10$^{-12}~$s$^{-1}$ at a radius $r = 0.028$~pc (or 5,800~AU).
Using Equation (\ref{dyn}), we estimate that a dynamical mass of 42~$\MO / (\sin i)^2$ is needed within this radius to gravitationally bind the rotating motion, where $i$ is the inclination of the rotation axis with respect to the line of sight. 

Our conclusion on the configuration of the bipolar outflow and the rotating core supports the original interpretation by \cite{lis93}. 
They noted the N-S linear velocity gradient in the core of $\sim$2 km s$^{-1}$ arcsec$^{-1}$ ($\omega = 1.7 \times$10$^{-12}~$s$^{-1}$ at the distance of 7.8 kpc), which is 40 percent smaller than our estimate above. 
This small discrepancy may be because of their larger synthesized beam (4$\arcsec$.5 $\times$ 3$\arcsec$.7 FWHM) and the fact that they measured the gradient in the N-S direction while the gradient is steepest at P.A. = 30$^\circ$.

The SO$_{2}$ emission in Figure \ref{pv}(b) exhibits a constant velocity gradient rather than the pattern of the Keplerian rotation that is characterized by an increase of the rotation velocity at closer distance to the central star. 
To further examine this point, we analyzed the SO$_{2}$ 29$_{4,26}$--28$_{5,23}$ (99.3925 GHz) and 32$_{5,27}$--31$_{6,26}$ (84.3209 GHz) emission that arise from energy levels at $E_{\rm{u}} =$ 441 K and 549 K from the ground state, respectively, much higher than $E_{\rm{u}} =$ 110.6 K for the 12$_{4,8}$--13$_{3,11}$ transition we have discussed so far. 
Both lines are free from contamination by other molecular lines, and are detected at the peak flux density of 0.44~Jy~beam$^{-1}$ (29$_{4,26}$--28$_{5,23}$) and 0.28~Jy~beam$^{-1}$ (32$_{5,27}$--31$_{6,26}$), with the distribution quite similar to each other. 
Figure \ref{pv}(d) shows the P-V diagram of the SO$_{2}$ 29$_{4,26}$--28$_{5,23}$ emission at P.A.=30$^\circ$ (color) in comparison with that of the SO$_{2}$ 12$_{4,8}$--13$_{3,11}$ emission (contours). 
Although the 29$_{4,26}$--28$_{5,23}$ emission arises from much hotter and denser molecular gas, the velocity gradient is similar to that of the 12$_{4,8}$--13$_{3,11}$ emission. 
This situation is quite different from the case of a disk in Keplerian rotation around a massive star, for which we should see a steeper velocity gradient for lines arising from the hotter and denser region closer to the star. 
We conclude that the ``line core" emission of SO$_{2}$ in Sgr\,B2(N), when observed at a $\sim$2$\arcsec$ 
(0.076 pc or 15,600 AU) resolution, arises from a rotating ring-like structure with a radius $\sim$6000 AU.

\subsection{Physical parameters of the bipolar outflow}

\subsubsection{Fractional abundances of SO$_{2}$ and SiO}

In order to estimate the physical parameters of the outflow, we need to know the fractional abundances of SO$_{2}$ and SiO,
defined as $X$(SO$_{2}$)=[SO$_{2}$]/[H$_{2}$] and $X$(SiO)=[SiO]/[H$_{2}$], 
which are known to vary significantly from an object to another \cite[e.g.,][]{mar92, cha97, van03, san12, san13}. 
We estimate the SO$_{2}$ and SiO fractional abundances by a comparison with the Sgr\,B2(N) outflow detected in C$^{18}$O, 
which is one of the molecular probes with stable fractional abundances.
Unfortunately, an analysis of the C$^{18}$O emission included in the present observations (Hasegawa et al., in prep.) 
shows that overlaps with the emission of HNCO ($v$$_{5}$=1), CH$_{3}$CN ($v$$_{4}$=1), and C$_{2}$H$_{5}$OCHO \cite[e.g.,][]{bel13}
lines make it difficult to analyze the high velocity C$^{18}$O wing emission over the full velocity range of the outflow. 
Instead, we choose a less-contaminated velocity channel at 46.6~km~s$^{-1}$ and compare with SO$_{2}$ and SiO.
Even at this velocity, point-like CH$_{3}$CN($v$$_{4}$=1) emission at K2 contaminates the C$^{18}$O image.
We avoid the K2 position and average C$^{18}$O, SO$_{2}$ and SiO emission over a 6$\arcsec$$\times$6$\arcsec$ box centered at 
($\alpha_{J2000}$, $\delta_{J2000}$)=[17$^{\mbox h}$47$^{\mbox m}$20$^{\mbox s}.$22, ~$-$~$\!\!$28$^{\circ}$22$\arcmin$19$\arcsec.$7], 
which is $\sim$4\arcsec\ away from the position of K2. 

For derivation of the SO$_{2}$ and SiO abundances, we assume 
that SO$_{2}$, SiO and C$^{18}$O lines are optically thin and in a local thermodynamical equilibrium 
(LTE, {\it i.e.,} they are excited to a common excitation temperature $T_{\rm ex} = T_{\rm ex}{\rm (C^{18}O)} = T_{\rm ex}{\rm (SO_2)} = T_{\rm ex}{\rm (SiO)}$). In addition, we assume that their relative abundances are uniform in the position-velocity space defined by the outflow.
For $T_{\rm ex}$, we have adopted a range of possible temperatures (50, 100, and 150~K) to compare the results.

Molecular column densities can be calculated \citep{liu98} from
\begin{eqnarray}\label{so2}
\label{massso2}
N &=& 2.04\times10^{20}\frac{{Q(T_{\rm{ex})}}\exp{(\frac{E_{u}}{T_{\rm{ex}}})}}
{{(\theta_{a}\times\theta_{b})}{\nu^{3}}{S\mu^{2}}}\int{S_{\nu}dv},
\end{eqnarray}
where $\theta_{a}\times\theta_{b}$ are the FWHM major and minor axes of the synthesized beam in units of arcseconds,  
$Q(T_{\rm{ex}})$ is the partition function, $E_{u}$ is the upper energy level in K, $\nu$ is the rest frequency of the transition in GHz,
and $S\mu^{2}$ is the product of the intrinsic line strength and the squared dipole momentum in D$^{2}$.
$S_{\nu}$ is the measured intensity in Jy~beam$^{-1}$, and $T_{\rm{ex}}$ is the excitation temperature in K. 
$\int{S_{\nu}dv}$ is the measured integrated line emission in Jy~beam$^{-1}$~km~s$^{-1}$.

For the SO$_{2}$ line, we use $Q$(150~K) = 2091, $Q$(100~K) = 1140, $Q$(50~K) = 404 from \cite{pic98}, $E_{u}$=110.6~K and $S\mu^{2}$=8.27~D$^{2}$ \citep{bel13}.
Similarly for the SiO line, we use $Q$(150~K) = 144, $Q$(100~K) = 96, $Q$(50~K) = 48 \citep{pic98}, $E_{u}$=6.25~K, $S\mu^{2}$=19.2~D$^{2}$ \citep{liu98, fer13}. 
The calculated column densities of SO$_{2}$ and SiO per unit velocity width at $V_{\rm LSR} = 46.6$~km~s$^{-1}$ are shown in Table \ref{ab} with the assumed excitation temperatures. 
For the fractional abundances, we derive the H$_{2}$ column densities per unit velocity width from the C$^{18}$O column densities per 
unit velocity width at $V_{\rm LSR} = 46.6$~km~s$^{-1}$ and the adopted $X$(C$^{18}$O) of $1\times10^{-7}$ for Sgr\,B2 \citep{lis89}.

Table \ref{ab} shows the SO$_{2}$ and SiO fractional abundances derived for $T_{\rm{ex}}$ = 50, 100, and 150~K. 
We note that the fractional abundances vary only mildly with the assumed LTE temperature, particularly for SiO. 
For the following discussion of the physical parameters of the outflow, we adopt the values derived for  $T_{\rm{ex}} = 100$~K with an uncertainty range 
estimated from the cases of $T_{\rm{ex}}$ = 50 and 150~K, {\it i.e.}, $X$(SO$_{2}$) = ${2.3^{+2.6}}_{-0.4}$$\times$10$^{-8}$ and $X$(SiO) = ${1.2^{+0.1}}_{-0.1}$$\times$10$^{-9}$.

These values are consistent with the previous estimates for other high-mass hot cores.  
\cite{van03}, \cite{esp13} derived molecular abundance of SO$_{2}$ 
in high-mass star-forming regions ranging from $X$(SO$_{2}$) = 10$^{-6}$ to 10$^{-8}$.
\cite{gus08}, \cite{ter11}, \cite{san13} and \cite{leu14} found the 
fractional abundance of SiO in the high-mass protostellar outflows ranging from $X$(SiO) = 10$^{-7}$ to 10$^{-9}$.
We should keep in mind that there is an uncertainty in our determination of the fractional abundances due to 
the assumption of uniform excitation and chemistry over space and velocity, 
which is the best allowed by the present data although it is obviously a bold simplification.

\subsubsection{Physical parameters}

\begin{figure}
\epsscale{0.65}
\plotone{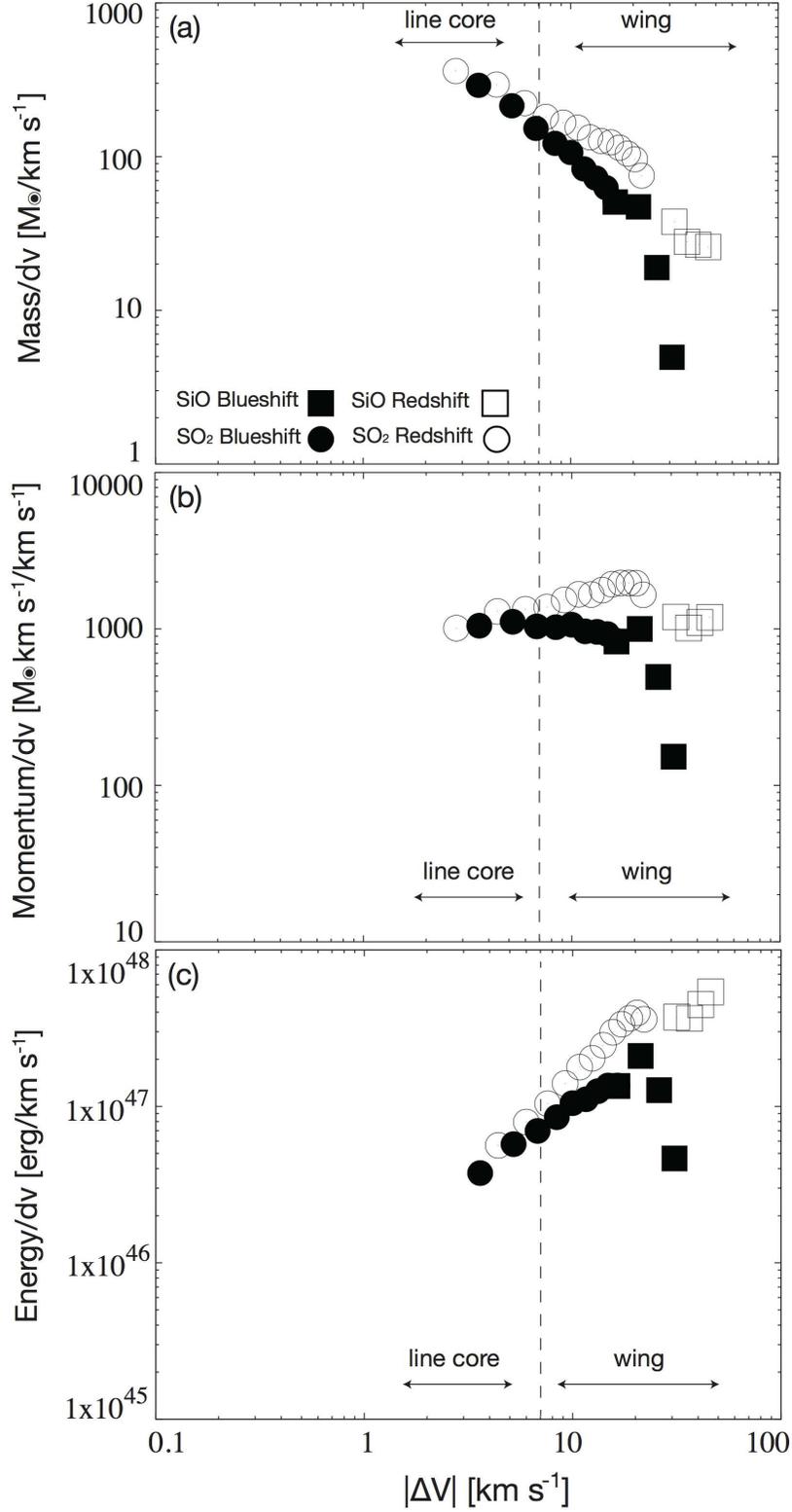}
\caption{
Masses, momenta, and energies of the outflow per unit velocity width as a function of velocity from the systemic velocity 
($|\Delta{V}|$=$V_{\rm{LSR}}$-$V_{\rm{SYS}}$) for the blueshifted and redshifted outflow lobes. The dashed line marks the 
defined separation between line core (low velocity) and wing (high velocity) emission.} 
\label{spectrum}
\end{figure}

Using the derived fractional abundances, we estimate the mass of the outflow material for individual velocity channels of the 
the SO$_{2}$ and SiO images.
Figure \ref{spectrum} shows the plots of (a) masses, (b) momenta, and (c) energies in the outflow per unit velocity width as 
a function of velocity offset from the systemic velocity ($|\Delta{V}|$=$|V_{\rm{LSR}}$-$V_{\rm{SYS}}|$).
The mass plot in Figure \ref{spectrum}(a) suggests an approximate power law in a form of $dM(v)$/$dv$$\propto$$|\Delta{V}|^{\gamma}$, 
with a power index $\gamma$$\sim$-1.
This slope is quite shallow compared with cases of outflows from low-mass protostars \cite[$\gamma \sim -4$,][]{she98}). 
\cite{ric00} and \cite{arc07} noted a tendency for the slope to steepen (from $\gamma \sim -1$ to $-10$) with the outflow age. 
The shallow slope of the Sgr\,B2(N) outflow is consistent with its youth ($t_{\rm dyn} \sim 5 \times 10^3$ years, see Table \ref{para}).

Figures \ref{spectrum}(b) and (c) show that the momentum is distributed rather evenly over the velocity range of the outflow,
and that the majority of the kinetic energy is carried at larger $|\Delta{V}|$.
It would be quite interesting to see how these plots compare with similar plots for other regions of high-mass star formation 
seen at high spatial resolution.

Table \ref{para} summarizes the physical properties of the molecular gas in the outflow. 
The momentum is given by $\Sigma$$M_{i}$$|\Delta{V_{i}}|$ and the energy by (1/2)$\Sigma$$M_{i}$${|\Delta{V_{i}}|}^{2}$, 
where $M_{i}$ is the outflow mass in the velocity channel $i$ and $\Delta{V}_{i}$ is its velocity offset relative to $V_{\rm{LSR}}$.
For the derivation of the outflow parameters, no correction for inclination angle was applied.  
We integrate only the velocity ranges of $|\Delta{V}|$ $>$ 7~km~s$^{-1}$ for deriving the physical parameters of the outflow.
The total mass of Sgr\,B2(N) outflow is derived as a sum of SiO emission 
($V_{\rm{LSR}}$= 32.2~km~s$^{-1}$ to 46.6~km~s$^{-1}$ for blueshifted velocities and 
$V_{\rm{LSR}}$= 89.8~km~s$^{-1}$ to 109~km~s$^{-1}$ for redshifted velocities) 
and SO$_{2}$ emission ($V_{\rm{LSR}}$=48.2~km~s$^{-1}$ to 54.6~km~s$^{-1}$ for blueshifted velocities and
$V_{\rm{LSR}}$= 72.2~km~s$^{-1}$ to 85~km~s$^{-1}$ for redshifted velocities).
Typical outflow velocity, $V_{\rm{outflow}}$ is derived from the ratio between momentum and mass. 
Comparing the spatial structure of the outflow with other massive outflows (e.g., W51 North, Zapata et al. 2009), the outflow in 
Sgr\,B2(N) has a relatively symmetric structure.

\cite{rid01} estimated the parameters of molecular outflows from 11 high-mass star-forming regions at a distance of $\sim$2~kpc.
In comparison with their physical parameters, the Sgr\,B2(N) outflow has a comparable mass ($\sim$2000~$\MO$), while
the flow size and the dynamical timescale of Sgr\,B2(N) is an order of magnitude smaller with the other high-mass star forming regions.
\cite{liu98} presented the SiO outflow from Sgr\,B2(M) whose mass of 100~$\MO$ assuming $X$(SiO) of 10$^{-7}$.
If they adopt $X$(SiO) of 10$^{-8}$ for Sgr\,B2(M), their outflow mass will be similar to the Sgr\,B2(N) result.
The dynamical timescales of G331.5--01 \citep{bro08,mer13} and W~51~North \citep{zap09} outflows are comparable to the Sgr\,B2(N) outflow.
The collimation of the Sgr\,B2(N) outflow is estimated from measuring FWHM of the peak emission of the outflow, 
which is $\sim$ 0.1~pc apart from the K2, and result in $\sim$~60$^\circ$.
From the comparisons, the Sgr\,B2(N) outflow can be characterized as a young and massive, but moderately collimated outflow.

In comparison with the outflows listed in \cite{zan01}, \cite{beu02}, the recompilation of \cite{wu04} and \cite{wu05}, 
the SgrB2(N) outflows with $\sim$2000~$\MO$ sits near the massive end of the spectrum. 
The Sgr\,B2(N) outflow can be a single massive outflow expected at the early stages of massive star formation \citep[e.g.,][]{zap10}, 
or alternatively the large mass can be a result of overlap and merge of the multiple outflows as in IRAS~16547--4247 \citep{hig15}. 
In order to resolve the outflow completely, observations with higher spatial resolution are needed.

\subsection{Possible Detection of SiO Maser Emission}\label{maser}

The maser emission of vibrationally excited SiO in regions of star formation is rare. 
It has been detected from only three regions of massive star formation so far, {\it i.e.,} Orion-KL \citep{sny74,tha74}, W51 IRS2 and Sgr\,B2(M) MD5 \citep{has86,uki87,mor92}, despite intensive searches for similar objects in star forming regions \citep{gen80,bar84,jew85,zap09b}.

\begin{figure}
\epsscale{0.7}
\plotone{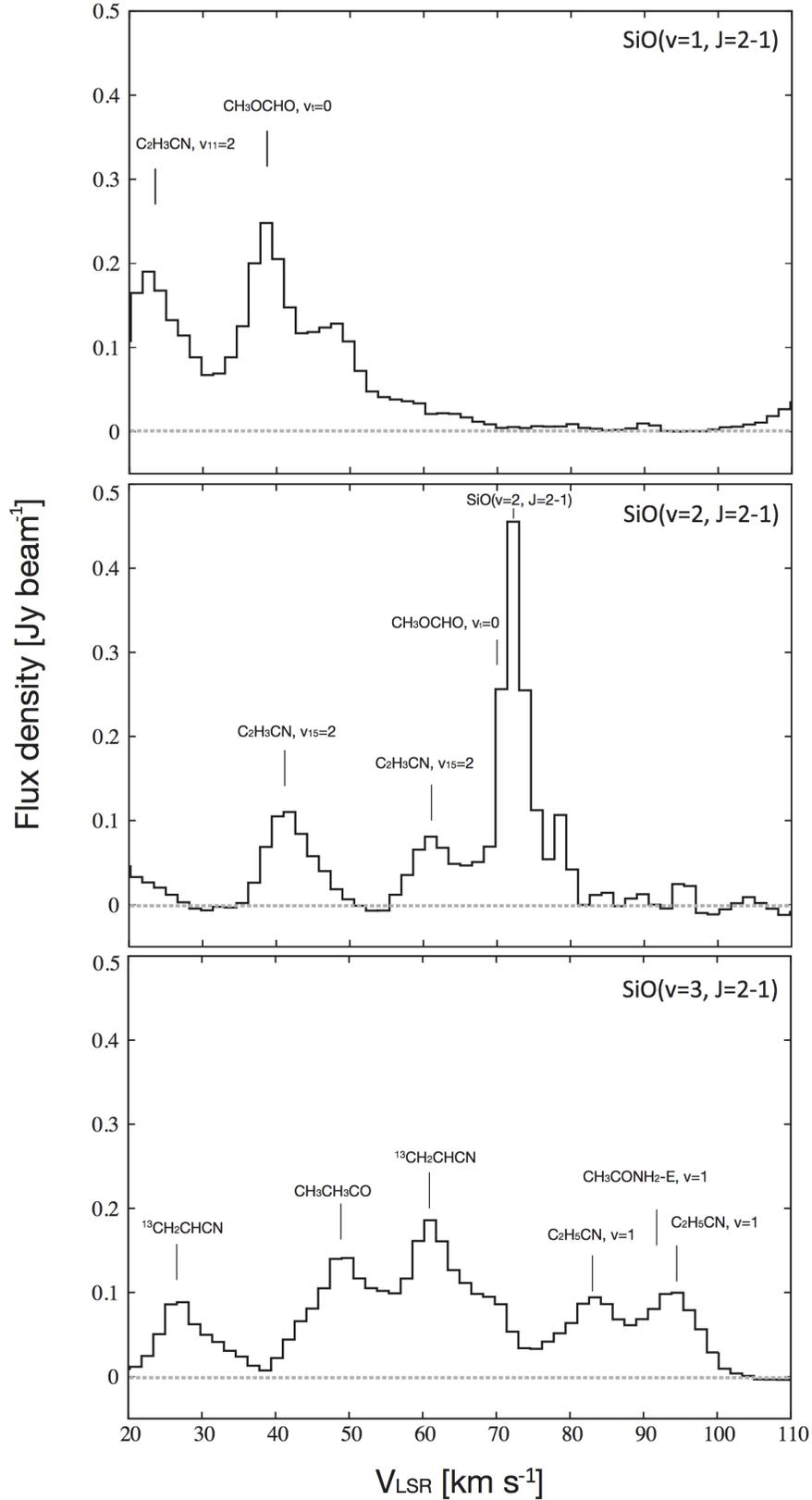}
\caption{Spectra of molecular lines from the K2 position of Sgr\,B2(N) around the frequencies of SiO ($v$=1,$J$=2--1) (top, undetected), SiO ($v$=2,$J$=2--1) (middle, tentatively assigned), and SiO ($v$=3, $J$=2--1) (bottom, undetected). The observations were made between August and October 2012 \citep{bel14}.}
\label{masers}
\end{figure}

We checked the ALMA data at the frequencies of the SiO ($J = 2 - 1$) lines in the $v$ = 1, 2, and 3 vibrationally excited states, and found an emission line for $v = 2$ with a peak flux density of 2~Jy at ${V}_{\rm{LSR}}$=72~km~s$^{-1}$ (Figure \ref{masers}).  
The emission has a linewidth of FWHM=3.8$\pm$0.3~km~s$^{-1}$, which is much narrower than the typical width of thermal emission lines from this region (FWHM$\sim$10~km~s$^{-1}$). 
Its spatial distribution is point-like (source size $\ll$ 1\farcs5) and its position, 
($\alpha_{J2000}$, $\delta_{J2000}$)=[17$^{\mbox h}$47$^{\mbox m}$19$^{\mbox s}.$86, ~$-$~$\!\!$28$^{\circ}$22$\arcmin$18$\arcsec.$5], 
is coincident with the position of K2 within 0\farcs2. 
The frequency of the detected emission corresponds to the CH$_{3}$OCHO ($v_{t} = 0, 4_{2, 3, 1} - 3_{1, 2, 2}$) line at ${V}_{\rm{LSR}}$ = 65~km~s$^{-1}$, but this assignment is unlikely because the emission in other transitions of this molecule have larger linewidth (FWHM$\sim$5~km~s$^{-1}$) and is spatially extended. 
The ${V}_{\rm{LSR}}$=72~km~s$^{-1}$ of the SiO ($v = 2, J = 2 - 1$) line emission is 9~km~s$^{-1}$ redshifted with respect to the systemic velocity of ${V}_{\rm{LSR}}$=63~km~s$^{-1}$, but it is within the velocity range of the H$_2$O masers. 
This kind of velocity offset is seen also in W51 IRS2 and Sgr B2(M) MD5 \citep{has86,zap09b}. 
From the total flux of the SiO ($v = 2, J = 2 - 1$) line emission, the isotropic photon luminosity is estimated as $L_{\nu }$ = 2.8 $\times$ 10$^{44}$~s$^{-1}$. 
This comfortably falls within the range of the SiO masers detected in other star forming regions \citep{zap09b}. 
No corresponding signal was found for the $v$=1 line exceeding 0.02~Jy(3~$\sigma$). 
The spectral range of the $v$=3 line overlaps with the lines from CH$_{2}$CH$^{13}$CN, 
CH$_{3}$CH$_{3}$CO, C$_{2}$H$_{5}$CN($v=1$) and CH$_{3}$C(O)NH$_{2}$, but there seems to be no SiO ($v$=3, $J$=2--1) line stronger than 0.4~Jy. 

Although SiO masers in various transitions in vibrationally excited states up to $v = 4$ are detected from many evolved stars such as Mira variables and red supergiants, the $v = 2, J = 2 - 1$ emission is anomalously weak, and exhibits some peculiarity when detected \citep{cla81,olo81,olo85,buj96,buj07}. 
\cite{olo85} has proposed that this behavior could arise from a population transfer due to line overlap between ro-vibrational transitions of SiO and H$_2$O in the masing regions around evolved stars. 
None of the three previously known SiO masers in star forming regions has been detected in the $v = 2, J = 2 - 1$ transition. 
In the case of the $J = 1 - 0$ SiO maser in W51 IRS2, only the $v = 2$ emission was detected during 1985 - 1989 with the $v = 1$ upper limits at $1/5$ to $1/15$ of the $v = 2$ intensities \citep{has86,fue89}. 
When \cite{zap09b} observed it in 2003, they found that the $v = 1, J=1 - 0$ emission was detectable at 1.0~Jy while the $v = 2$ emission had become weaker at 2.5~Jy.
As our understanding of the excitation mechanism for the SiO masers in star forming regions is still limited, we cannot rule out the maser assignment from the fact that only the $v = 2, J = 2 - 1$ transition is detected. 

The characteristics of the emission line described above, {\it i.e.,} the point-like spatial distribution and extremely narrow linewidth, 
make it quite probable that the line has a maser nature. 
It is important to confirm its assignment with the SiO ($v = 2, J = 2 - 1$) line by, {\it e.g.,} observing the matching $J = 1 - 0$ lines, measuring polarization, or setting a high enough lower limit to the line brightness temperature with a higher spatial resolution. 
If confirmed, it will be not only the 4$^{\rm th}$ star forming region with SiO maser emission, but also the first such object with the $v = 2, J = 2 - 1$ emission. 
The SiO maser emission will provide crucial information on the structure, kinematics and physical condition of the close vicinity (within $\sim 100$ AU) of the massive protostar that drives the outflow.

\subsection{Insight on the Process of Massive Star Formation}\label{msf}

Based on the observations of a sample of hot molecular cores around massive (proto-) stars, 
\cite{ces06} and 
\cite{belt11} have proposed that the rotating cores are classified into two classes, {\it i.e.,} circumstellar disks and circumcluster toroids. 
In their scenario of massive star formation, a larger scale infalling envelope provides the mass to the toroid, which is a transitional structure that feeds the mass towards the accretion disks with Keplerian rotation around individual forming stars in the central cluster. 
Recent observations with millimeter and submillimeter interferometers provide increasing evidences of disk-like structures of radii 1000 - 2000 AU with signatures of Keplerian rotation around B or O type (proto) stars \citep[e.g.,][]{ces14,hun14,belt14,joh15}. 
Compared with these cases, the rotating SO$_{2}$ core we found around K2 in Sgr\,B2(N) may fall in the class of toroids, 
because it is a much larger ring-like structure without the signature of the Keplerian rotation. 

The structure of Sgr\,B2(N) presented here shows an intriguing resemblance to that of W51 North. 
Both objects are embedded in very luminous regions of massive star formation, and have the system of bipolar outflow and a large rotating hot core \citep[e.g.,][]{zap09,zap10}. 
The bipolar outflow of W51 North is well traced by the SiO (5 - 4) and CO (2 - 1) emission. 
Its mass is $\sim 200 \MO$, which is an order of magnitude less than what we find in Sgr\,B2(N), but is still a massive outflow \citep[see, e.g.,][]{zan01,beu02,wu04,wu05}. 
Near the driving source is a cluster of luminous H$_2$O masers that span a large velocity range, and a SiO maser \citep[e.g.,][]{mor92,eis02}. 
At the center is a peak of millimeter and submillimeter dust emission and a hot rotating core surrounds it \citep[][]{zap09,zap10}. 
The hot core appears as a rotating toroid with a central cavity 3000 AU in radius, when observed in SO$_2$ ($22_{2,20} - 22_{1,21}$) and other molecular lines arising from energy levels 20 - 800 K above the ground state. 
Molecular emission from even higher energy levels are more spatially confined and fills in the cavity with the kinematics reproduced by a model of a Keplerian disk with an infalling motion. 
Based on these observations, \cite{zap10} proposed a possible evolutionary sequence of massive star formation in four phases, in which W51 North is placed in Phase II, the large and massive pseudo-disk with layers of physical conditions. 
Sgr\,B2(N) may be in the same phase or a little later, with a continuum source K2 detected at centimeter wavelengths. 

At this point, the structure and kinematics of the molecular gas in Sgr\,B2(N) inside the SO$_2$ toroid is not known. 
Is it a single and coherent structure as postulated in the {\it core accretion} model of massive star formation, or, alternatively, a cluster of forming stars with accretion disks collectively contributing to the large luminosity and the massive bipolar outflow in the scenario of {\it competitive accretion} \citep[e.g.,][]{tan14}?
The modest spatial resolution ($\sim$2$\arcsec$) of the ALMA data we analyzed here leaves this important question open. 
Further ALMA observations with higher spatial resolution would answer the question and uncover the processes that link the Sgr\,B2(N) toroid to the massive star formation inside it. 

\section{SUMMARY}
We have analyzed the archival data of the ALMA observations of Sgr\,B2(N) in SO$_{2}$($12_{4,8} - 13_{3,11}$) and SiO ($2 - 1$) 
lines at an angular resolution of $\sim$2$\arcsec$ to investigate the kinematic structure of the region. 
Our main findings are summarized as follows:

\begin{enumerate}
\item 
Sgr\,B2(N) has a system of a bipolar outflow and a rotating core, as originally interpreted by \cite{lis93}. 
The SiO line shows the bipolar outflow whose driving source is located at K2, 
and the SO$_{2}$ line shows both the outflowing gas and the rotating hot core.
We note that H$_{2}$O masers are associated with innermost part of the bipolar outflow.

\item
Sgr\,B2(N) outflow is characterized as a young (5$\times$10$^{3}$~yr) and massive ($\sim$2000~$\MO$) outflow. 
It is moderately collimated ($\sim$60$^{\circ}$).

\item Fractional abundances of SO$_{2}$ and SiO in the outflowing gas are estimated by a comparison with 
the C$^{18}$O ($1-0$) emission to be $X$(SO$_{2}$) of ${2.3^{+2.6}}_{-0.4}$$\times$10$^{-8}$ and $X$(SiO) of 
${1.2^{+0.1}}_{-0.1}$$\times$10$^{-9}$ for the assumed excitation temperature of $100 \pm 50$~K.
These values are consistent with the estimates in other high-mass star forming regions. 

\item 
The mass spectrum of the outflow suggests an approximate power law in a form of $dM(v)$/$dv$$\propto$$|\Delta{V}|^{\gamma}$, 
with a power index $\gamma \sim -1$.
The shallow slope is consistent with the youth of the Sgr\,B2(N) outflow.
The outflow momentum is distributed rather evenly over the velocity range of the outflow,
while a majority of the kinetic energy is carried at larger $|\Delta{V}|$.

\item 
We discovered a point source of narrow line emission at the position of K2 in Sgr\,B2(N) that can possibly 
be assigned to the SiO ($v = 2, J = 2 - 1$) maser emission, although this assignment needs a confirmation. 
When confirmed, this will make Sgr\,B2(N) the 4$^{\rm th}$ star forming region with detected SiO masers.

\item 
The hot rotating core found in the SO$_2$ emission has a ring-like structure with a radius $\sim 6000$ AU without a 
clear sign of Keplerian rotation, and it falls in the class of {\it toroid} classified by \cite{belt11}. 
Sgr\,B2(N) exhibits striking resemblance to W51 North, although we do not know the structure inside the toroid of Sgr\,B2(N). 
Compared with W51 North, Sgr\,B2(N) may be in the same or a little later phase in the scenario of massive star formation proposed by \cite{zap10}.

\end{enumerate}
Although the overly rich emission lines detected from Sgr\,B2(N) require careful selection of molecules and transitions for proper analyses, 
ALMA data are proven to be very useful in understanding the kinematics and physical parameters of this high-mass star formation region. 
Sgr\,B2 complex is one of the most important regions as sources for understanding high-mass star formation in ALMA-era.

\bigskip
We thank the anonymous referee for careful reading and constructive comments that helped greatly to improve the manuscript.
We also thank the ALMA staff for the observations during the commissioning stage. 
This letter makes use of the following ALMA data: ADS/JAO.ALMA$\#$2011.0.0017.S. 
ALMA is a partnership of ESO (representing its member states), NSF (USA), and NINS (Japan), 
together with NRC (Canada), NSC, and ASIAA (Taiwan), in cooperation with the Republic of Chile. 
The Joint ALMA Observatory is operated by ESO, AUI/NRAO, and NAOJ.
Data analysis were carried out on common use data analysis computer system at the Astronomy Data Center, 
ADC, of the National Astronomical Observatory of Japan.
We acknowledge Takashi Tsukaghi and Koichiro Sugiyama for their contributions.

\appendix

\begin{deluxetable}{l l l l l l l c c c c c c}
\tabletypesize{\small}
\rotate
\tablecaption{Parameters for the ALMA observations of Sgr\,B2(N)\label{obs}}
\tablewidth{0pt}
\tablehead{
\colhead{Molecule} & \colhead{Transition} & \colhead{$\nu$}& \colhead{Synthesized Beam} & \colhead{Velocity resolution} & \colhead{rms noise level} \\
\colhead{} & \colhead{} & \colhead{[GHz]}& \colhead{} & \colhead{[km~s$^{-1}$]} & \colhead{[mJy~beam$^{-1}$]}}
\startdata
 SO$_{2}$ & 12$_{4,8}$--13$_{3,11}$ & 107.843 & 1\farcs9$\times$1\farcs4 & 1.6 & 6 \\
 SO$_{2}$ & 29$_{4,26}$--28$_{5,23}$ & 99.393 & 1\farcs9$\times$1\farcs4 & 1.6 & 6 \\
 SO$_{2}$ & 32$_{5,27}$--31$_{6,26}$ & 84.321 & 2\farcs2$\times$1\farcs6 & 1.6 & 7 \\
 SiO & $v$=0, $J$=2--1 & 86.847 &  2\farcs5$\times$1\farcs8 & 1.6 (spectrum) and 4.8 (channel maps) &  7 (channel maps) \\
 SiO & $v$=1, $J$=2--1 & 86.243 &  2\farcs5$\times$1\farcs8 & 1.6 &  3 \\
 SiO & $v$=2, $J$=2--1 & 85.640 &  2\farcs5$\times$1\farcs8 & 1.6 &  3 \\
 SiO & $v$=3, $J$=2--1 & 85.038 &  2\farcs5$\times$1\farcs8 & 1.6 &  3 \\
\hline
\enddata
\tablecomments{The rms noise level is derived in the emission-free area defined by inspecting the channel maps.}
\end{deluxetable}

\begin{deluxetable}{l l l l l c c c c}
\tablecaption{Column densities and fractional abundances in the outflowing gas$^a$ \label{ab}}
\tablewidth{0pt}
\tablehead{
\colhead{} & \colhead{$T_{\rm{ex}}$=50~K} & \colhead{$T_{\rm{ex}}$=100~K}& \colhead{$T_{\rm{ex}}$=150~K}}
\startdata
$N_{\rm{SO_{2}}}/dv$ [cm$^{-2}$/km~s$^{-1}$]$^b$ & 8.4$\times$10$^{14}$ & 7.8$\times$10$^{14}$ & 9.9$\times$10$^{14}$ \\
$N_{\rm{SiO}}/dv$ [cm$^{-2}$/km~s$^{-1}$]$^b$ & 2.2$\times$10$^{13}$ & 4.2$\times$10$^{13}$ &  6.2$\times$10$^{13}$ \\
$N_{\rm{C^{18}O}}/dv$ [cm$^{-2}$/km~s$^{-1}$]$^c$ & 1.7$\times$10$^{15}$ & 3.5$\times$10$^{15}$ & 5.2$\times$10$^{15}$\\
$N_{\rm{H_{2}}}/dv$ [cm$^{-2}$/km~s$^{-1}$]$^d$ & 1.7$\times$10$^{22}$ & 3.5$\times$10$^{22}$ & 5.2$\times$10$^{22}$\\
\hline
$X$(SO$_{2}$) & 4.9$\times$10$^{-8}$  & 2.3$\times$10$^{-8}$  & 1.9$\times$10$^{-8}$ \\
$X$(SiO) & 1.3$\times$10$^{-9}$ & 1.2$\times$10$^{-9}$ & 1.2$\times$10$^{-9}$ \\
\hline
\enddata
\tablecomments{$^a$Measured at $V_{\rm LSR} = 46.6~$km~s$^{-1}$ by averaging over a 6$\arcsec$$\times$6$\arcsec$ box centered at ($\alpha_{J2000}$, $\delta_{J2000}$)=[17$^{\mbox h}$47$^{\mbox m}$20$^{\mbox s}.$22, ~$-$~$\!\!$28$^{\circ}$22$\arcmin$19$\arcsec.$7].
$^b$Present work. $^c$Hasegawa {\it et al.}, in preparation. $^d$From $N_{\rm{C^{18}O}}/dv$ and $X$(C$^{18}$O) = $1\times10^{-7}$ \citep{lis89}. }
\end{deluxetable}

\begin{deluxetable}{l l l c c}
\tabletypesize{\small}
\tablecaption{Physical parameters of the outflow \label{para}}
\tablewidth{0pt}
\tablehead{
\colhead{Parameter} & \colhead{}}
\startdata
 Distance from K2 to SiO peak position\tablenotemark{a}: $l_{1}$ [pc] \\ 
 ~~~Blue lobe & 0.08 \\
 ~~~Red lobe & 0.06 \\
 Distance from K2 to SiO outer contour\tablenotemark{b}: $l_{2}$ [pc] \\
 ~~~Blue lobe & 0.3 \\
 ~~~Red lobe & 0.2 \\
 Outflow mass ($\MO$)  \\
 ~~~Blue lobe\tablenotemark{c} & 5.7$\times$10$^{2}$ \\
 ~~~Red lobe\tablenotemark{d}  & 1.4$\times$10$^{3}$ \\
 ~~~Total & 2.0$\times$10$^{3}$ \\
 Momentum ($\MO$~km~s$^{-1}$)  \\
 ~~~Blue lobe & 7.4$\times$10$^{3}$ \\
 ~~~Red lobe  & 2.2$\times$10$^{4}$ \\
 ~~~Total & 3.0$\times$10$^{4}$ \\
 Kinetic energy (erg)  \\
 ~~~Blue lobe & 1.1$\times$10$^{48}$ \\
 ~~~Red lobe  & 4.4$\times$10$^{48}$ \\
 ~~~Total & 5.5$\times$10$^{48}$ \\
 Typical outflow velocity\tablenotemark{e} : $V_{\rm{outflow}}$ (~km~s$^{-1}$) \\
 ~~~Blue lobe &  13 \\
 ~~~Red lobe & 16 \\
 ~~~Total & 15 \\
  Dynamical time\tablenotemark{f} : $t_{\rm{dyn}}$ (yr) \\
 ~~~Blue lobe & 5.6$\times$10$^{3}$ \\
 ~~~Red lobe & 3.7$\times$10$^{3}$ \\
 ~~~Average & 4.7$\times$10$^{3}$ \\
\hline 
\enddata
\tablecomments{This table shows the physical parameters of the bipolar outflow traced by SiO and SO$_{2}$; 
size, mass, momentum, kinetic energy, outflow velocity, and dynamical time.}
\tablenotetext{a}{$l_{1}$ is measured as the SiO peak from K2 position from Figure \ref{outflow}.}
\tablenotetext{b}{$l_{2}$ is measured as the farthest 5~$\sigma$ contour from K2 position from Figure \ref{outflow}.}
\tablenotetext{c}{Mass, momentum, and kinetic energy of blueshifted components are derived from integration in the velocity range of $|\Delta{V}|$ $>$ 7~km~s$^{-1}$ ($V_{\rm{LSR}}$= 32.2~km~s$^{-1}$ to 46.6~km~s$^{-1}$ for SiO and $V_{\rm{LSR}}$=48.2~km~s$^{-1}$ to 54.6~km~s$^{-1}$ for SO$_{2}$).}
\tablenotetext{d}{Mass, momentum, and kinetic energy of redshifted components are derived from integration in the velocity range of $|\Delta{V}|$ $>$ 7~km~s$^{-1}$ 
($V_{\rm{LSR}}$= 72.2~km~s$^{-1}$ to 85~km~s$^{-1}$ for SO$_{2}$ and $V_{\rm{LSR}}$= 89.8~km~s$^{-1}$ to 109~km~s$^{-1}$ for SiO).}
\tablenotetext{e}{$V_{\rm{outflow}}$ is derived from the ratio between momentum and mass.}
\tablenotetext{f}{$t_{\rm{dyn}}$ is calculated from the ratio between ${l_{1}}$ and $V_{\rm{outflow}}$ for the blueshifted and redshifted components.}
\end{deluxetable}

\begin{figure}
\epsscale{0.9}
\plotone{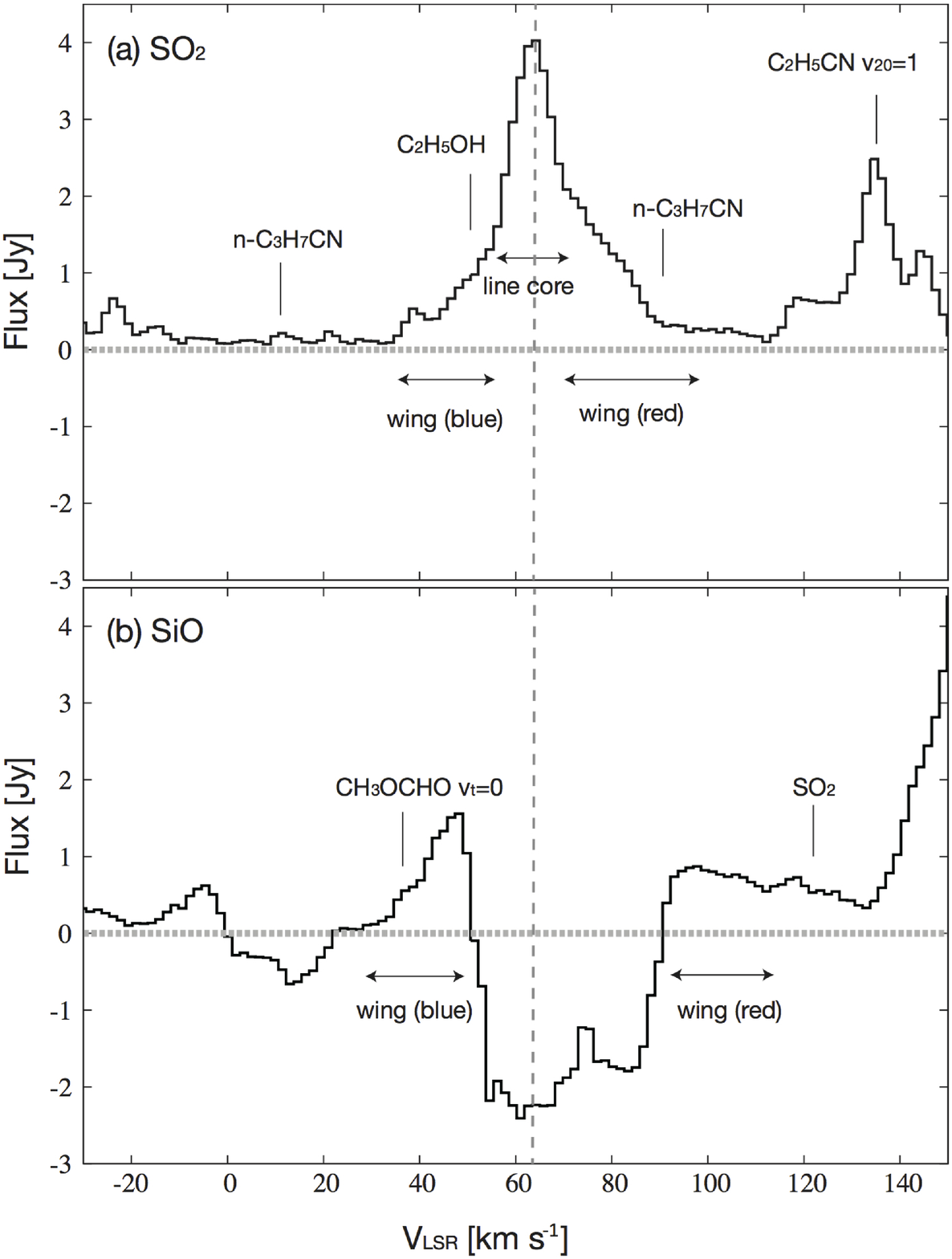}
\caption{
Integrated spectra around the (a) SO$_{2}$ and (b) SiO emission lines in Sgr\,B2(N).  
The spectra were integrated over a region of 20$\arcsec$$\times$20$\arcsec$ region, centered on the continuum source K2.
$V$$_{\rm{LSR}}$ of 63~km~s$^{-1}$ is the systemic velocity of Sgr\,B2(N).}
\label{prof}
\end{figure}

\end{document}